# *Edges in Brain Networks: Contributions to Models of Structure and Function*

Joshua Faskowitz[1,2], Richard F. Betzel[1,2,3], and Olaf Sporns[1,2,3]

1. Program in Neuroscience, Indiana University, Bloomington, IN, USA
2. Department of Psychological and Brain Sciences, Indiana University, Bloomington, IN, USA
3. Indiana University Network Science Institute, Indiana University, Bloomington, IN, USA

## Abstract

Network models describe the brain as sets of nodes and edges that represent its distributed organization. So far, most discoveries in network neuroscience have prioritized insights that highlight distinct groupings and specialized functional contributions of network nodes. Importantly, these functional contributions are determined and expressed by the web of their interrelationships, formed by network edges. Here, we underscore the important contributions made by brain network edges for understanding distributed brain organization. Different types of edges represent different types of relationships, including connectivity and similarity among nodes. Adopting a specific definition of edges can fundamentally alter how we analyze and interpret a brain network. Furthermore, edges can associate into collectives and higher-order arrangements, describe time series, and form edge communities that provide insights into brain network topology complementary to the traditional node-centric perspective. Focusing on the edges, and the higher-order or dynamic information they can provide, discloses previously underappreciated aspects of structural and functional network organization.

## Introduction

Modern neuroscience has come to appreciate the complexity of the brain's wiring structure and functional dynamics. Increasingly, neuroscientists employ the tools of network science to model the brain as a network, a mathematical representation of data well suited to investigate complex systems (Bullmore and Sporns 2009, Bassett and Sporns 2017). Brain networks can reveal many aspects of brain structure and function, including clusters and modules (Betzel, Medaglia et al. 2018), or information flow and communication (Avena-Koenigsberger, Misic et al. 2018). Approaching the brain as a network, a connectome (Sporns, Tononi et al. 2005) composed of distinct elements and their interrelationships, naturally integrates local and global perspectives, linking the roles of individual network elements to distributed function. In essence, networks map neural architecture from neurons, neuronal populations and large-scale regions, to their mutual relationships.

There are many ways to map and represent connectomes. For a select few "model" organisms, the micro-scale, single neuron networks of the compete nervous system have been meticulously documented *via* electron microscopy (White, Southgate et al. 1986). Other approaches, using techniques that afford less spatial resolution while offering broader coverage, have yielded meso and macroscale connectomes across many species, including humans. For example, noninvasive imaging allows the brain to be represented as a network of inferred paths of axonal tracts through the white matter (Hagmann, Cammoun et al. 2008), of morphometric similarity between parts of the cortex (Seidlitz, Vasa et al. 2018), or of functional correlation of intrinsic hemodynamic fluctuations across time (Biswal, Mennes et al. 2010). Brain networks provide a universal modeling framework enabling comparisons across data modality, scale, and species.

The nodes of brain networks are generally taken to represent distinct neural elements, such as neurons, neuronal populations, or regions, while the edges record the dyadic (pairwise) relationships between these elements. Fundamentally, these two components of the network model are inseparable. Nodes would not connect without edges, and edges would be nonsensical without nodes. Yet, when applied to the brain, network models often prioritize nodes, describing and differentiating their mutual relations and functional contributions. Examples of key 'node-centric' concepts are highly connected and central hubs, which integrate information, or densely connected, coherent communities of nodes associated with specialized functional systems. Furthermore, networks are often globally described through distributions of measures like node degree, strength, clustering, or participation coefficient, and the network's community structure is almost exclusively expressed as nodal partitions. Finally, node metrics are frequently used to probe for associations with behavioral or genetic traits. Less heralded are the edges—while providing crucial information to make these nodal network assessments, they are rarely seen as primary descriptors of network organization. The focus on the nodal characteristics extends prevailing trends in the long history of brain mapping, which has been dominated by the search for localized neural elements that relate to specific functions (Raichle 2009).

Even though edges are half of the network model, many issues concerning the brain's interrelationships have so far been underappreciated. The edges of the brain, and their collective topology, are key ingredients that transform and elevate static maps of the brain ("wiring diagrams") into distributed and dynamic systems capable of supporting behavior and cognition. Here we shine a spotlight on brain network edges, surveying the ways in which information located between the nodes can be used to understand brain network organization. We begin by clarifying that the type of edge, supported by underlying neural data, is consequential for the downstream network analyses. Then, we review the various constructs that edges can jointly form, which are useful because they can capture relationships that extend beyond pairwise interactions. We cover the importance of edges for studying brain communication and briefly review ways in which communication dynamics evolve over time at the edge level. Finally, we look to the future, and include a discussion of several new developments for interpreting information at the edge level. Overall, we endeavor to bring attention to the importance of brain network edges, and to demonstrate the value in carefully considering the information they provide.

# Network Construction

Networks offer a universal language to describe complex systems made up of many interacting parts. The basic ingredients for any network are its nodes and edges. The nodes describe the discrete elements of a system, whereas the edges express the relationships that can be measured between these elements. While the definition of networks as sets of nodes and edges is universal, which real-world constructs are taken to be nodes and which as edges depends on assumptions and interpretations that guide the construction of the network model (Butts 2009). Depending on the system being modeled, edges may be binary or may carry a weight. Weights may be both positive and negative, and they may express directed or undirected relations. In many real-world networks, like a social network, the subway map, or a power grid, these basic network ingredients are generally well-defined and accessible to data collection. In contrast, defining the nodes and edges of a brain network is less straightforward.

Aside from the micro-scale, where it could be argued that nodes and edges unambiguously correspond to neurons and synaptic contacts (Fig. 1a), representing brain data as a network requires choosing from a wide range of node definitions as well as picking a valid mode and metric for their interrelationships (Bassett, Zurn et al. 2018). As such, it has been demonstrated that definition of nodes and nodal parcellations can significantly influence the results of downstream network analyses (Zalesky, Fornito et al. 2010, Arslan, Ktena et al. 2018, Messe 2020). Edge definition is just as consequential. Focusing on the brain's interrelationships, we can broadly classify edges as documenting connectivity or similarity between the brain's nodes. Additionally, edges can be annotated with supplemental measurements or carry weights that reflect the fusion of multiple modalities (Box: Alternative Weighting Strategies).

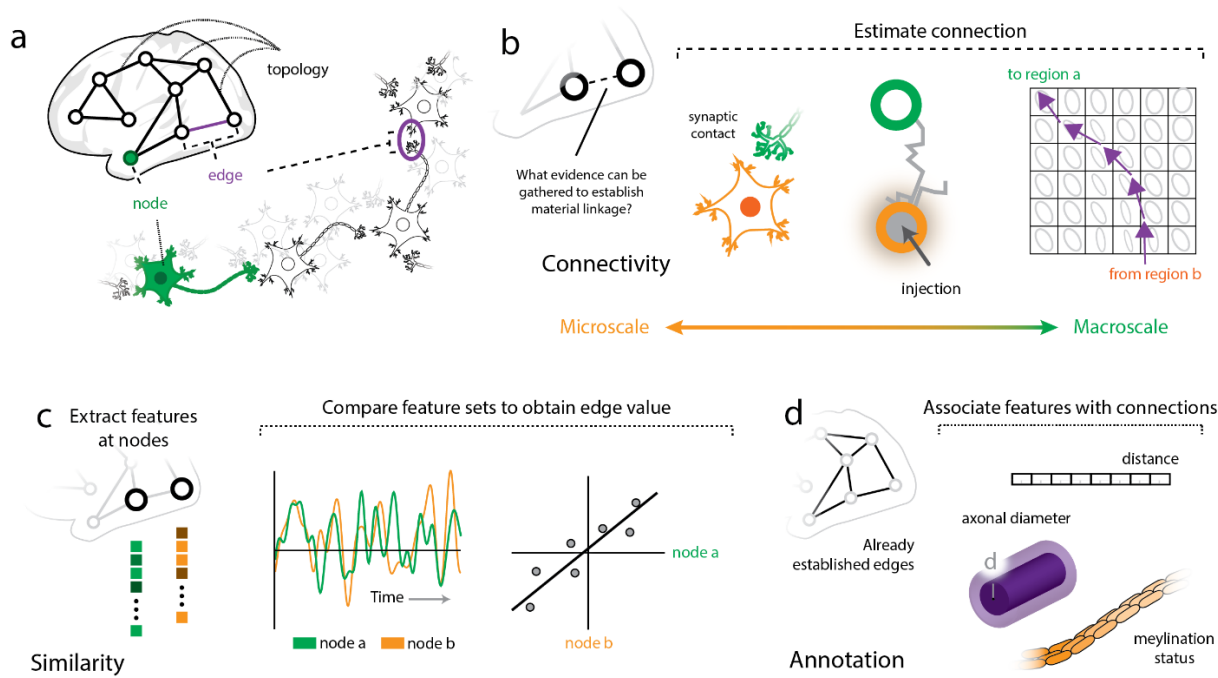

**Figure 1** The relational content of the brain can be documented in several manners; **a**) The basic components of a brain network, the *nodes* and *edges*, can be modeled across scales, spanning neurons to cortical regions; **b**) Edges of connectivity report the ways in which nodes can be materially linked, across spatial scales; at the microscale, these edges can represent neuronal contact whereas at the macroscale, such edges can be estimated via computational processes like tractography; **c**) Edges of similarity report the ways in which feature sets at nodes are alike; such features sets can be gathered from both dynamic and static data; **d**) Edges can be annotated with weights from other modalities or embeddings, adding an additional layer of information on the network.

## Edge Types: Connectivity

Edges can represent connectivity between neural elements, quantifying material linkage or contact, supporting information flow, signal spread, or communication, and summarized in a sparse connectivity or adjacency matrix (Fig. 1b). Depending on data modality, connectivity can be resolved from the micro- (White, Southgate et al. 1986) to the macroscale (Hagmann, Cammoun et al. 2008), providing varying levels of evidence of a true (physical) connection. At the microscale, edges represent synapses or gap junctions, resolved with techniques such as electron microscopy or through light-microscopic labeling and imaging (Motta, Berning et al. 2019). At increasing scales, neural data documents coarser patterns of connectivity that link populations of neurons comprising one or more cell types or layers or representing entire brain regions. In mammalian brains, such interregional connections are often myelinated, collectively forming the brain's white matter, and can be mapped with a variety of techniques. For example, tract tracing is used to label and reconstruct interregional projections (Markov, Ercsey-Ravasz et al. 2014, Oh, Harris et al. 2014). Generally, multiple reconstructions need to be combined to achieve robust characterization of connection patterns and weights. One approach is to informatically collate the literature of tract tracing experiments, to create comprehensive maps that also record ordinal assessments of connection weights (Kotter 2004, Bota, Sporns et al. 2015). At the scale of millimeters, bundles of topographically organized axonal paths through the white matter, commonly referred to as tracts, can be estimated via tractography (Jbabdi, Sotiropoulos et al. 2015) and serve to quantify connectivity (Sotiropoulos and Zalesky 2019, Yeh, Jones et al. 2020). Common to these edge definitions expressing connectivity is a notion of anatomical substrate enabling various patterns of between-node communication. A different approach aims to infer patterns of effective connectivity that correspond to causal relationships and influences (Friston 2011). Effective connectivity is estimated from functional data via methods that establish statistical or model-based causality between time-varying nodal signals (Valdes-Sosa, Roebroeck et al. 2011, Reid, Headley et al. 2019) or perturbational evidence (Lim, Mohajerani et al. 2012). Ultimately, edges of connectivity define the potential for one node to influence another, made possible by estimated anatomical linkage.

### Edge Types: Similarity

Edges can also denote the similarity between node-level features (Fig. 1c). Computing the statistical similarity (or distance) between each pair of nodal feature sets forms a dense similarity matrix (all entries are nonzero), which may be interpreted as a network. Notably, the feature sets at each node reflect datapoints collected across space or time, which modulates the interpretation of such edges. Using imaging or histological observations, neuroanatomical features can be sampled at each node, including for example cortical thickness (Carmon, Heege et al. 2020) or layer intensity profile (Paquola, Vos De Wael et al. 2019). These features can then be statistically compared within or across-subjects (Alexander-Bloch, Giedd et al. 2013) to create edges that represent the similarity of feature sets. The interpretation of similarity-based edges varies depending on what is included in the feature set. The strength of such anatomical similarity edges can be interpreted as shared developmental or genetic influence. For instance, structural similarity, which may reflect cytoarchitectonic similarity, is thought to relate to anatomical connectivity (Goulas, Majka et al. 2019). Another similarity-based approach quantifies correlated gene expression between areas of cortex (Richiardi, Altmann et al. 2015), made possible by extensive brain atlases documenting genetic profiles in stereotaxic space (Ng, Bernard et al. 2009, Hawrylycz, Lein et al. 2012). Edges based on correlated gene expression among a set of genes known to be enriched in supra-granular cortex align with canonical system organization (Krienen, Yeo et al. 2016) and show significant association with edges of structural covariance (Romero-Garcia, Whitaker et al. 2018). Finally, the informatic collation of functional activation experiments provides across-study evidence that certain region pairs co-activate more readily than others, forming meta-analytic co-activation edges (Crossley, Mechelli et al. 2013).

Recordings of activity timeseries at neural elements may be taken to represent temporally resolved feature sets whose similarity, or more generally, statistical association, is widely employed to interrogate brain organization. Neural activity can be recorded across a range of resolutions and frequencies, and in turn, can serve as the basis of many types of bivariate similarity calculations (Smith, Miller et al. 2011, see also Basti, Nili et al. 2020). Neural recordings with high temporal precision, such as electrical potentials or magnetic fields (Hari and Puce 2017), provide data allowing the resolution of directed, non-linear, and/or information theoretic edge weights (Astolfi, Cincotti et al. 2007, Ince, Giordano et al. 2017). Brain signals recorded at lower temporal resolution, such as the blood oxygen level dependent (BOLD) signal or $Ca^{2+}$ recordings, can be compared using Pearson correlation or wavelet coherence. Such edges are generally referred to as "functional connectivity" (Friston 2011), essentially encapsulating the collective node dynamics in the form of a covariance matrix (Reid, Headley et al. 2019). A looming topic in studies of functional connectivity is that of the dynamics of functional relationships, and if observed fluctuations in similarity represent neurobiologically relevant processes or mere statistical variance in an otherwise stationary relationship (Laumann, Snyder et al. 2017, Lurie, Kessler et al. 2020). Relatedly, the similarity of dynamics could be influenced by cognitive state, raising the question whether the recorded edge represents a trait or state measurement (Geerligs, Rubinov et al. 2015). Dynamics at each node can also be used to collect large feature sets of time series properties (Fulcher and Jones 2017), which can be used to compare temporal profile similarity (Shafiei, Markello et al. 2020), an edge measure that is distinct from correlation and can reveal dynamical hierarchies.

## Edge-Centric Network Analyses

Once a brain network is constructed, common practice is to use the tools of network science and graph theory to describe the organizational patterns of the data (Rubinov and Sporns 2010, Fornito, Zalesky et al. 2016). In many instances, network analyses are used to obtain information about nodes, asking questions like: Which nodes are most influential, or highly connected? How can these nodes be meaningfully grouped?

Network analyses that result in information at the edge level provide complementary insights. A common edge construct is the path, an ordered sequence of unique edges that links a source to a target node. Edgewise metrics based on paths include the edge betweenness centrality which describes the fraction of shortest paths that traverse a specific edge. Paths are important for network communication as they define possible routes for signal and information flow. Communication models use network paths ("routing") or random walks ("diffusion") to estimate the potential for communication between nodes, resulting in a dense communication matrix where each edge expresses a valuation of this potential (Goni, van den Heuvel et al. 2014, Seguin, Tian et al. 2020). Finally, the vulnerability of networks can be assessed by removing network components and observing the resulting effect (Henry, Duffy et al. 2020). Removal of specific edges allows, for example, to record effects on global network statistics (de Reus, Saenger et al. 2014, Ardesch, Scholtens et al. 2019). This "edge-lesioning" approach can be applied to a range of common network measures, including those that produce measurements per node like clustering coefficient, and hence can assess the global effect of edge removal.

Network science also offers approaches to represent a *network of edges*, to focus on how the edges relate to each other. One approach is to construct a line graph which documents how edges share nodes. Whereas a traditional network documents adjacency, or how nodes are linked via edges, a line graph documents incidence, or how edges are linked via common nodes (Evans and Lambiotte 2009). For the line graph network representation, the network is essentially flipped inside-out, with edges from the original network becoming nodes. In practice, the line graph has matrix dimensions of *E*-by-*E*, where *E* is the number of unique edges of the original network. A notable property of line graphs is that high degree nodes (hubs) in the original network become dense incidence clusters (cliques) in the line graph. Networks of edges distinct from line graphs can also be obtained by computing edge-similarity matrices. For example, an *E*-by-*E* similarity matrix may be obtained using the Jaccard index applied to edges (Ahn, Bagrow et al. 2010). Clustering such edge similarity matrices, or any *E*-by-*E* matrix, results in edge communities. These communities give rise to overlap at the level of nodes, where each node can be affiliated with multiple communities assigned to its emanating edges. Clustering a line graph of structural connectivity reveals bilateral spatially coherent link communities, with differential connectivity scores per community, and community overlap that converges on nodes that are traditionally considered hubs (de Reus, Saenger et al. 2014).

Networks are a universal phenomenon, and generally, the algorithms we apply to networks to uncover clustered, community, or scale-free organization are data agnostic. This means that network measures like the clustering coefficient are easy to compute on a power grid, a brain network, or any other sort of network in hand with a minimal set of assumptions (fulfilling the requirements of a *simple graph*, a network without self-loops and hyperedges). However, while it is possible to run the gamut of network tools on brain data, doing so without considering the source of the neural data and the ensuing interpretation of nodes and edges is unwise. The incorporation of domain-specific neuroscience expertise—knowledge about the neural data source, and an understanding of how a network measure relates to the aspect of brain organization being modeled—should be a key consideration when analyzing brain networks.

Edges in brain networks can be defined in different ways. Importantly, information about how an edge was constructed and the underlying relationship that the edge is intended to represent affects how the network should be analyzed. Take for example path-based measurements applied to brain networks. Paths over structural edges are intuitive and have physical meaning, given that a path may represent hypothetical signal propagation over a material substrate (Mišić, Betzel et al. 2015, Avena-Koenigsberger, Mišić et al. 2017). For such structural paths, its constituent edges and edge weights should reflect the cost or capacity of communication between nodes, such as distance, speed, volume, or bandwidth.

Paths over functional edges that express similarity are less intuitive, and possibly ill-conceived, compared to paths over edges of connectivity. What does a path over functional similarity measurements mean? One possible argument is that structural and functional edge weights are indeed positively associated (Honey, Sporns et al. 2009), so that paths over functional similarities may, to some extent, be associated with underlying connectivity. However, given that measures such as Pearson's correlation express mixtures of direct and indirect sources of variance in a networked setting (Zalesky, Fornito et al. 2012, Sanchez-Romero and Cole 2021), this interpretation is likely too charitable. Another approach for using functional edges to construct paths is to study the transient routes that appear along the underlying structural graph (Griffa, Ricaud et al. 2017). Network paths and their derived measures should be interpreted differently based on edge type, as they likely capture different organizational features of a brain network.

Another instance in which the edge definition influences network analysis is the case of surrogate data modeling, when an empirical network measurement needs to be compared to hypothetical, yet plausible, network topologies. Null or generative models should be able to create surrogate data that recapitulates certain network characteristics, but with a different pattern of edges (Betzel, Avena-Koenigsberger et al. 2016, Rubinov 2016, Faskowitz and Sporns 2020). Such null models are important, for example, for the application of modularity maximization, which searches for clustered edge weights above a baseline rate commonly estimated with an edge-swapping null model. However, for brain networks constructed from statistical comparisons, there exist more suitable null models that account for signed edges (Rubinov and Sporns 2011) or spatial information (Esfahlani, Bertolero et al. 2020) and take into account the transitive relationships between edges (Zalesky, Fornito et al. 2012). In applications of community detection and beyond, null models that account for the physical distance distribution of edges are a more accurate model of the brain, which is spatially embedded (Horvat, Gamanut et al. 2016, Roberts, Perry et al. 2016) (see Box: Spatial Embedding Makes Brain Networks Unique). Surrogate data that does not account for the distance distribution of edges will be less efficiently embedded, with longer connections than expected (Bassett, Greenfield et al. 2010). Network science offers a range of null and generative models which neuroscientists can choose from or modify, to better align with edge definition.

Many observable real-world networks are sparse, in that relatively few edges exist out of all the possible pairwise node combinations. Estimates of structural connectivity between nodes are also observed to be sparse, particularly at finer spatial resolution and greater distances, possibly an outcome of selection pressure on wiring cost (Bullmore and Sporns 2012). In contrast, similarity assessments result in fully dense networks that present practical and conceptual challenges for network analyses. Some practitioners may opt to selectively remove edges below a threshold to enforce sparsity (Garrison, Scheinost et al. 2015, Fallani, Latora et al. 2017), with thresholds chosen according to across-group consensus (van den Heuvel, de Lange et al. 2017, Betzel, Griffa et al. 2019) or to retain a network feature such as a connected component or minimum spanning tree (Tewarie, van Dellen et al. 2015, Nicolini, Forcellini et al. 2020). Thresholding can induce biases and confounds (Zalesky, Fornito et al. 2012) in the overall network topology and therefore must be performed with justification and with an understanding that different thresholds could possibly affect the investigation's main findings. Alternatively, analytical approaches that incorporate noisy edges or imperfect graph observation could be a fruitful future direction for network neuroscience (Young, Cantwell et al. 2020).

## Edge Constructs: From Motifs to Higher-Order Relations

Edges on their own report a straightforward relational quantity. These quantities can be treated as elementary network features, to be associated with traits and behaviors through mass univariate testing, in what is sometimes referred to as a bag-of-edges approach or brain-wide association (Chung, Bridgeford et al. 2021). However, edges may also be grouped together to form richer constructs that capture distributed patterns of brain organization. Small groups of edges form constructs that can be analyzed as building blocks or primitives of the complete network. Mass univariate methods could fail to uncover these higher-order relationships, and even prove to be underpowered (Zalesky, Fornito et al. 2010), because they focus on edges as independent entities. Here we describe edge-based constructs moving from more localized patterns such as motifs or connectivity fingerprints to more global patterns of brain network topology.

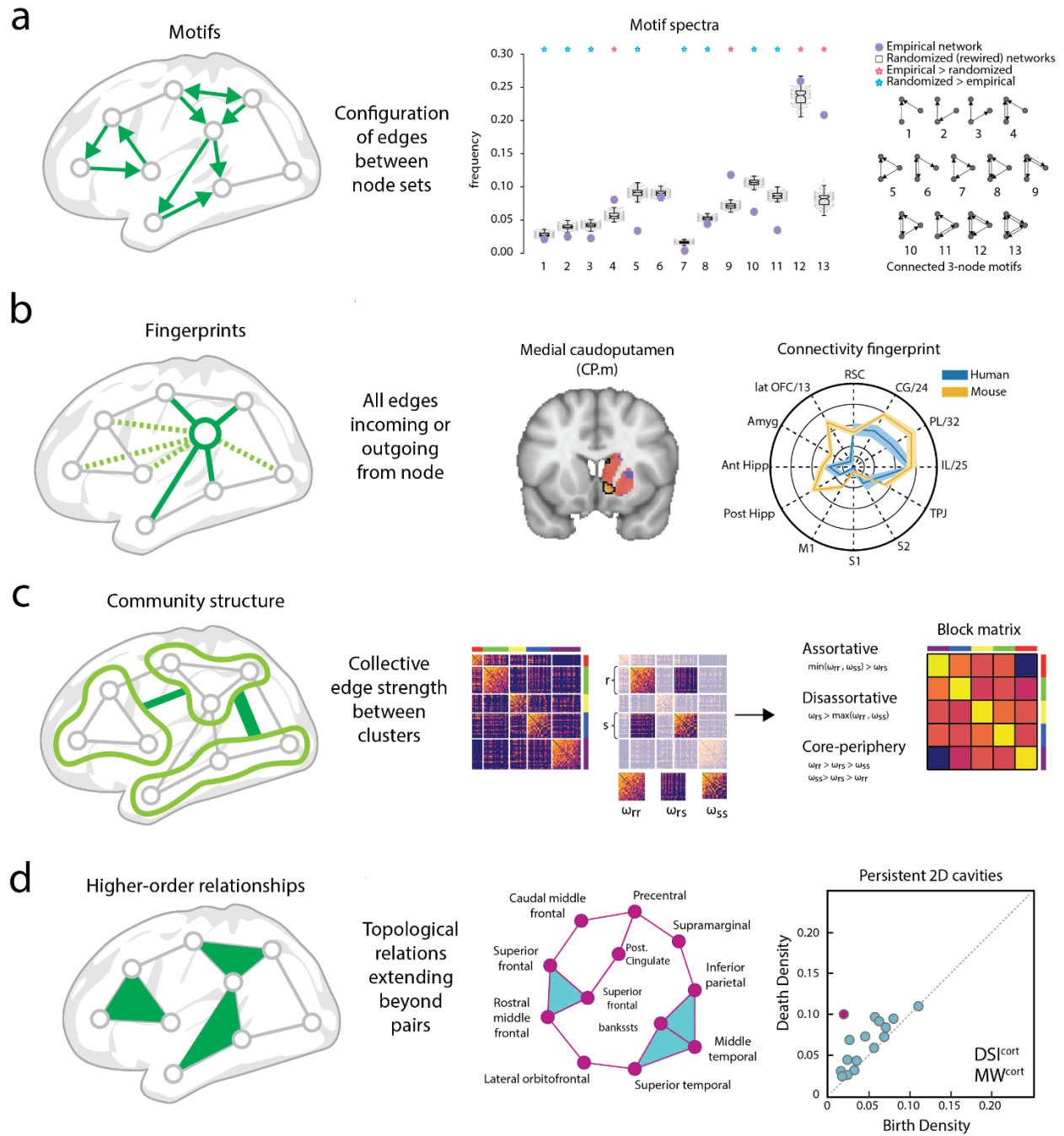

**Figure 2** Edges can be grouped to form constructs amenable for analysis; **a**) Motifs are characterized by a set number of nodes and the pattern of edges that fall between them; the motif spectra visualizes the frequency of various motifs present in the network; figure adapted from Liu, Zheng et al. (2020); **b**) Connectivity fingerprints describe the set of edges connected to a specific node, which can create a global context or profile for a specific region and can be used to identify homologs across species; figure adapted from Balsters, Zerbi et al. (2020); **c**) Community structure describes a mesoscale organization of the network, which can be used to calculate and classify edge strengths between clusters; figure adapted from Betzel, Medaglia et al. (2018); **d**) Higher-order relationships, such as cliques and cavities, can be built by aggregating pairwise relationships to assess higher dimensional structure of the network; figure adapted from Sizemore, Giusti et al. (2018).

## Motifs

Network motifs are subgraphs with a fixed number of nodes and differentiated by the pattern of edges falling between these nodes (Fig. 2a). For example, between three connected nodes, there are 13 topologically unique ways that edges (directed and unweighted) can be placed, forming 13 motifs (Fig 2a). The frequency of that each motif's expression tells us about the network's local building blocks (Sporns and Kotter 2004, Dechery and MacLean 2018). Motif frequencies are assessed using surrogate networks, to gauge the under- or over-expression of certain motifs (Horvat, Gamanut et al. 2016, Liu, Zheng et al. 2020) or can be related to principal dimensions of network organization (Morgan, Achard et al. 2018). The edge configurations of specific motifs constrain the possible patterns of dynamic interactions (Sporns and Kotter 2004) and enable temporal coherence and synchrony. For example, motif configurations containing bi-directional connections, termed resonance pairs, can induce zero-lag synchrony in a variety of neuronal spiking models, despite non-zero conduction delays on individual edges (Gollo, Mirasso et al. 2014). Taken together, network motifs express intermediate aspects of brain architecture and are thus informative for investigating how the wider network might support functional activity.

## Connectional Fingerprints

In virtually all brain networks, the pattern of incoming and outgoing edges attached to each node is unique. These edge patterns, known as connectional fingerprints (Fig. 2b), were proposed as fundamental structural profiles that shape the functional specialization of a given region by determining from whom that region receives its inputs and to whom its outputs are delivered (Passingham, Stephan et al. 2002, Mars, Passingham et al. 2018). The fingerprinting approach can help to clarify the functional roles regions might play, based on their differential weights to other areas (Tang, Jbabdi et al. 2019), or to predict functional activation patterns (Osher, Saxe et al. 2016, Saygin, Osher et al. 2016). A key concept of the fingerprinting approach is the embedding of areas within an abstract connectivity space, as opposed to a geometric space (Mars, Passingham et al. 2018). The connectivity space can be used, in conjunction with common structures, to help identify homologies between species (Balsters, Zerbi et al. 2020). Furthermore, this connectivity space can be used to subdivide larger regions based on fine-grained connectivity profiles (Genon, Reid et al. 2018).

From a network perspective, a connectivity fingerprint is a row or column of the adjacency matrix which records a vector of edge weights attached to each node. Notably, this row of edge weights is a discrete analogue of traditional seed-based connectivity. The similarity of edge patterns can be measured using the normalized matching index (Fornito, Zalesky et al. 2016) or cosine similarity (Betzel and Bassett 2018), to gauge connectional homophily between nodes, which is a critical ingredient for generative models of brain networks (Betzel, Avena-Koenigsberger et al. 2016). Ultimately, the pattern of edges emanating from each node describes the context of the node within the larger network architecture. The connectivity fingerprinting approach demonstrates the utility of assessing a complete pattern of connections to each node, rather than looking at only a subset.

## Community Structure

Although network communities are often interpreted from a node-centric perspective—most commonly defined as groupings of densely connected nodes—it is the edges that determine which nodes should be grouped together, whether by strength of connection (Sporns and Betzel 2016) or by similarity of edge connectivity patterns (Moyer, Gutman et al. 2015, Faskowitz, Yan et al. 2018). Given an established or inferred community structure, the edges that fall between communities are used to characterize the integrative hub-like roles of select nodes. For example, edge information is used to identify nodes whose edges are highly dispersed amongst functional areas (Bertolero, Yeo et al. 2015) or to classify hub areas associated with different cognitive domains (Gordon, Lynch et al. 2018). Furthermore, the community structure can be used to reduce the network to its block structure, by recording the summed or averaged edge strength between communities (Fig. 2c). This block structure characterizes meso-scale between-community connection patterns, such as modular, core-periphery, or disassortative configurations (Betzel, Medaglia et al. 2018, Faskowitz and Sporns 2020).

## Higher-Order Relationships

Thus far, we have reviewed the ways groups of edges form constructs that can be used to probe the organization of a brain network. Groups of edges can capture patterns beyond the pairwise relationship reported by a single edge (Fig. 2d). Another avenue for uncovering such patterns is to employ the tools of algebraic topology (Battiston, Cencetti et al. 2020), which provide a formal mathematical framework for analyzing the higher-order relational content of a network using concepts such as cliques and cavities (Giusti, Pastalkova et al. 2015, Sizemore, Phillips-Cremins et al. 2019). Applied to brain data, such tools show how all-to-all components of a network may serve to localize hub-like roles that some brain areas might play (Sizemore, Giusti et al. 2018) or help to elucidate spiking activity progression in large neuronal microcircuit simulations (Nolte, Gal et al. 2020). An advantage of these approaches is the ability to describe how components of the ordinary network of pairwise relationships take part in higher-order mesoscale organization, observable by applying mathematical reformulations like filtrations. Applications have highlighted the increase in integrative organization after administration of psychoactive drugs like psilocybin by identifying edges that support topological cycles (Petri, Expert et al. 2014). Algebraic topology also offers new ways to draw relationships between nodes based on clustering in a low-dimensional embedding space (Patania, Selvaggi et al. 2019).

Without edges, a network would merely be a set of nodes with no relational content. All network assessments, even the ones that produce node-wise measurements like clustering coefficient, need edge data. Evidently, edges are trivially important for network analysis. This section highlighted the further utility of edge groupings to understand levels of organization in brain networks. These approaches complement other methods like psychophysiological interaction analysis (O'Reilly, Woolrich et al. 2012) or bundle analysis (Chandio, Risacher et al. 2020) which provide ways to extract rich multivariate data about inter-areal relationships outside of a network context. Overall, the complex structural and functional organization of the brain can be explored through relational information. In particular, the features that form from groups of edges, from motifs to fingerprints to communities and cliques establish local relationships that enable specific functional capabilities or place nodes within a global connectivity context.

## Edges in Communication and Brain Dynamics

The history of neuroscience provides us with vast cumulative knowledge about the localization of structural and functional features across the cortex and subcortex, from the micro to the macro scale, resulting in comprehensive maps of the brain (Amunts and Zilles 2015, Poldrack and Yarkoni 2016). Through extensive brain mapping studies, specific areas can be associated with specialized function, tuned to a behavior or cognitive processes. Such maps document the spatial layout of areas, but not necessarily how these areas interact. The addition of edges to a map provides information about how the elements of a map collectively form an integrative system, supportive of both local and distributed activity. Edges are also key for studying brain communication. They can represent the structural scaffold on which communication unfolds and channel the ongoing dynamic activity between neural elements (Avena-Koenigsberger, Misic et al. 2018). Here we examine the role of edges, and information at the edges, for understanding how the brain forms an integrative communicating system.

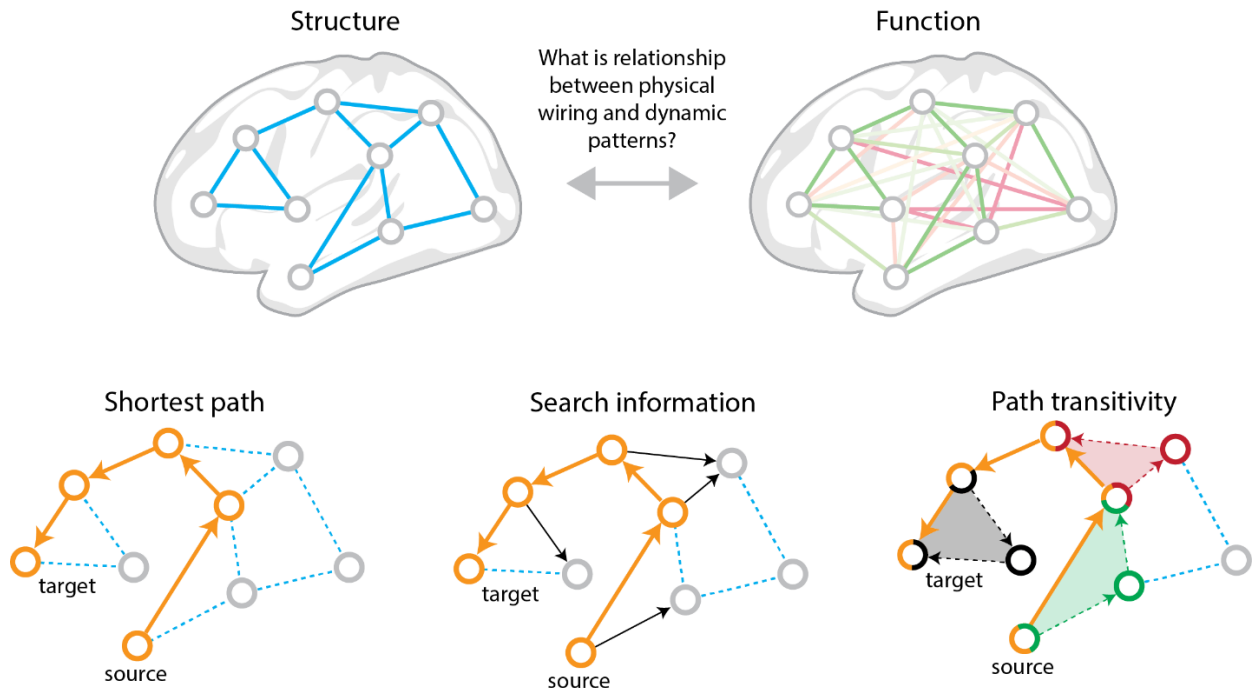

**Figure 3** Edges can report both anatomical and functional relationships between regions; how these two topologies relate to each other remains an important topic of investigation for network neuroscience; one way to approach this question is to model how communication processes, guided by certain algorithmic rules, might unfold over the structural edges. In shortest paths routing, communication between a source and target node unfolds along the shortest structural path. From the perspective of a diffusion process or knowledgeless random walker, accessing the shortest path may be difficult if there exist opportunities for the walker to "hop" off the path (we show these opportunities as black arrows in the middle panel). The total information (usually expressed in units of bits) required to navigate the shortest path successfully is referred to as "search information." Even if a random walker diverges from the shortest path, there may be opportunities to return. This intuition is quantified by the measure "path transitivity", which identifies cases where, following deviations from a network's shortest path, a random walker can return. In the above figure, we highlight three such cases, plotting the deviation and return as a filled triangle (red, black, and green).

## Structure-Function Relationships

A profitable starting point for investigating brain communication is to assess the relationship between structural and functional network organization (Bansal, Nakuci et al. 2018, Suarez, Markello et al. 2020), to observe the extent to which structural edge weights estimated *in vivo* possibly constrain the resultant functional topology. Focusing on edge weights, we can find a moderate positive association between structure and function at group and individual levels in humans (Honey, Sporns et al. 2009, Zimmermann, Griffiths et al. 2018), across node sets (Messe 2020), and even in other species including invertebrates (Turner, Mann et al. 2021). However, the structure-function relationship is more complex than implied by an edge-wise comparison -- for example, it can be confounded by overlap and transitivity (Zalesky, Fornito et al. 2012) and biased by distance (Honey, Sporns et al. 2009). Notably, the communication that takes place between network nodes is a complex mixture of effects due to numerous intersecting paths (Avena-Koenigsberger, Mišić et al. 2017). The observed statistical dependence at any one edge is a result of communication through direct connections and a mix of local and global contexts.

Thus, structure-function relationships may be better modeled by utilizing information beyond the pairwise connectivity. Take for example, the comparison of structural and functional connectivity fingerprint coupling at each node (Vazquez-Rodriguez, Suarez et al. 2019, Baum, Cui et al. 2020), which follow smooth gradients of functional topography. Other sorts of higher order contexts, such as embedding vectors generated from biased random walks of the network (Rosenthal, Váša et al. 2018, Levakov, Faskowitz et al. 2021), can predict the functional topology with greater accuracy.

Since structural edges may provide a scaffold on which communication takes place (Fig. 3), it makes sense that network communication modeling has been taken up by neuroscientists to explain structure-function relationships. Many communication models are based on network paths over a topology that is assumed to be efficiently wired, based on metabolic and volumetric constraints. Communication models based on paths taken over the structural topology produce edgewise information about the ease of communication between nodes, e.g., diffusion (Abdelnour, Voss et al. 2014), search information (Goni, van den Heuvel et al. 2014), and navigability (Seguin, van den Heuvel et al. 2018, Vázquez-Rodriguez, Liu et al. 2020). These coefficients, or combinations thereof, can predict (or correlate with) the functional topology. The incorporation of higher-order information, or polysynaptic signaling, not only improves alignment with the empirical functional topology, but also increases the predictive utility of structural connectivity, allowing for better prediction of broad behavioral dimensions (Seguin, Tian et al. 2020).

Understanding the mapping from structure to function has been scrutinized using frameworks ranging from communication modeling (Avena-Koenigsberger, Misic et al. 2018) to deep learning (Sarwar, Tian et al. 2021) to neural mass modeling (Sanz-Leon, Knock et al. 2015). In this pursuit, the target goal is made more difficult by the fact that most pairwise estimates of dynamic interaction, communication or functional connectivity are averaged over time. Time-averaged estimates of functional similarity could be insensitive to important dynamics at the edge level that reflect communication processes. Therein lies a motivation for observing edgewise and time-resolved functional connectivity.

## Time-varying Functional Connectivity

We expect that communication between brain regions would ebb and flow over short time scales, reflected in a sequence of correlation or coupling values at each edge. These dynamics could be in response to varying cognitive demands and environmental cues or reflect a dynamic repertoire of intrinsic functionality. Recent emphasis has been placed on tracking and quantifying how functional coactivation changes moment-by-moment between nodes, termed dynamic or time-varying functional connectivity (Lurie, Kessler et al. 2020). In practice, time-varying connectivity resolves the transient relationships between regions, which can signal different internal states that the brain is occupying or passing through (Fukushima, Betzel et al. 2018). These dynamics are driven by external stimuli (Simony, Honey et al. 2016), and are associated with clinical grouping or outcome (Douw, van Dellen et al. 2019), or patterns of structural topology (Shen, Hutchison et al. 2015, Zamora-Lopez, Chen et al. 2016, Fukushima and Sporns 2020).

There are two main approaches for studying time-varying connectivity, using either model-based dynamical systems that simulate the activity of neural populations, or data-driven statistical evaluations that operate on the observed timeseries (Lurie, Kessler et al. 2020). A common data-driven method for rendering dynamic correlation values is by subdividing the empirical timeseries into many overlapping windows. For each window, a correlation matrix is calculated, generating a sequence of values at each edge representing changing co-activity from window to window. Such an approach is subject to key parameter choices, like window length and offset (Shakil, Lee et al. 2016) that can affect the detection of potentially blur sharp or instantaneous periods of synchrony.

### Edge Time Series

Recently, a new approach has been proposed that obviates the need for sliding windows, while recovering a frame-by-frame account of an edge's activity (Faskowitz, Esfahlani et al. 2020, Zamani Esfahlani, Jo et al. 2020). An edge time series is constructed by multiplying the z-scored signals of two nodes, which also happens to be an intermediate step of calculating Pearson's correlation (van Oort, Mennes et al. 2018). These time series track each edge's functional co-fluctuations at the same temporal resolution as the original signal. Applying this construct to fMRI data, we observe intermittent high amplitude "events" of co-fluctuation that account for a large portion of the classic time-averaged functional connectivity. This finding implies that the time-averaged FC estimate is driven by brief epochs of burst-like activity (Tagliazucchi, Balenzuela et al. 2012, Liu and Duyn 2013, Thompson and Fransson 2016). Interestingly, high amplitude frames reflect a shared functional organization, and yet, also exhibit deviations to reliably distinguish subjects from each other (Betzel, Cutts et al. 2021). A further property of edge time series is that, at any given frame, the instantaneous co-fluctuation pattern is partitioned into exactly two communities (Sporns, Faskowitz et al. 2021). This feature implies that canonical functional systems are only transiently expressed, and that their familiar brain-wide architecture results from the superposition of many bipartitions over time.

By recovering temporally resolved time series for each edge, the communication dynamics can be studied with high precision. The simple Pearson correlation "unwrapping" procedures can readily be extended to domains beyond fMRI such as electrophysiological recordings. Such recordings afford much higher sampling rates and could be analyzed with a variant of the edge time series that adds lag terms and hence could possibly establish directionality of the edge dynamics. In a further extension, at the neuronal level, models of spike transmission at the edge (synapse) level can be built (McKenzie, Huszar et al. 2021). Additionally, mutual information can be "unwrapped" into pointwise mutual information that can also record time-resolved edge fluctuations (Lizier 2014). Findings based on edge time series complement previous map-based approaches (Liu and Duyn 2013), which also focus on the co-fluctuating activity at single frames. There remains much to be explored regarding the networked edge dynamics, including the ongoing topology these dynamics form (Betzel, Cutts et al. 2021) and the co-fluctuation patterns that might evolve intrinsically (Lindquist, Xu et al. 2014) or evoked during experimental manipulations (Rosenthal, Sporns et al. 2017, Cooper, Kurkela et al. 2021).

# Future Directions

## Relationships Between Edges

The common conceptualization of brain networks follows a familiar formula, which we have reviewed here, with $N$ nodes describing the physical neural elements and the $E$ edges describing the web of various types of interrelationships between these elements (Fig 4a). In this approach, we take the neural elements to be the fundamental units, to be compared in a pairwise manner (but see Box: Parcellating Nodes or Edges). An alternative approach would be to take the *edges* as the fundamental units (Ahn, Bagrow et al. 2010), to construct edge-edge matrices that index the similarity between edge information, particularly over time (Bassett, Wymbs et al. 2014, Davison, Schlesinger et al. 2015, see also Iraji, Calhoun et al. 2016, Faskowitz, Esfahlani et al. 2020).

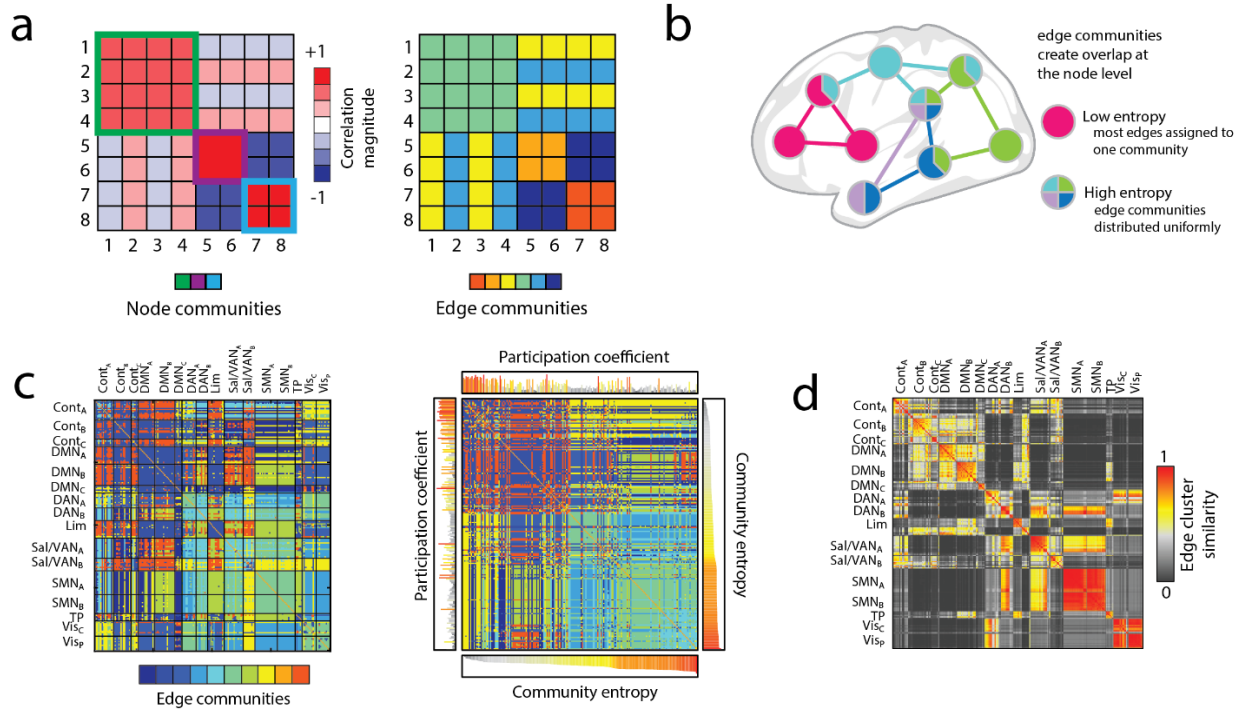

**Figure 4** Edge-centric approaches allow for edges to be clustered directly, which can reveal mesoscale organization at the edge level; **a**) Node-based clustering results in groupings of nodes that are commonly modular, and can be visualized as dense squares on the diagonal of an adjacency matrix; edge-based clustering results in groupings of edges with a common property, and can be visualized by coloring the adjacency matric with community affiliation; **b**) Edge community overlap can be indexed by a node-level measurement of entropy, which characterizes the distribution of discrete communities connecting to each node; **c**) A 10-community clustering of edge functional connectivity visualized as an adjacency matrix (left), and sorted by community entropy (right); the sorted matrix displays a 'banding' pattern, which demonstrates a difference between high- and low-entropy nodes; figure adapted from Faskowitz, Esfahlani et al. (2020); **d**) The edge cluster similarity matrix indicates the similarity of edge community profiles, which are rows (or columns) of the edges community matrix (as in **c**); this matrix indicates the varying levels of edge community diversity contained within canonical functional systems; figure adapted from Faskowitz, Esfahlani et al. (2020).

Comparing the pairwise temporal co-fluctuation profiles of edges enables the creation of hyperedges, to reveal temporally similar edge bundles that evolved in a task-specific manner (Davison, Schlesinger et al. 2015). These profiles can also serve as the basis of inter-subject dynamic similarity evaluated during a movie watching task, which can flow between integrated and segregated topologies related to stimulus properties (Betzel, Byrge et al. 2020) or serve as the basis to investigate higher-order correlations related to narrative content (Owen, Chang et al. 2019). Comparing edge time series in a pairwise fashion results in an edge functional connectivity (eFC) matrix (Faskowitz, Esfahlani et al. 2020). Clustering this matrix exposes a pervasively overlapping community structure (Fig 4b-d) at the node level that not only bridges canonical systems, but also reveals nested edge-level structure for diverse canonical systems like the control and default mode network (Jo, Esfahlani et al. 2020). Edge functional connectivity may also contain new sources of individual variation (Jo, Faskowitz et al. 2020). Taken together, these approaches suggest that taking the edges as fundamental network components provides a new perspective through which to interrogate brain organization.

## White Matter Matters

The white matter is the anatomical tissue that, by volume, comprises over half of the human brain. In terms of inter-areal connectivity, the *white matter matters* (Fields 2008). The dogma that the white matter is 'passive wiring' is being challenged by evidence that the myelin plays a role in how action potentials are propagated through the brain, which in turn could affect oscillatory activity in the cortex (Fields, Woo et al. 2015). At a macroscopic level, lesions in the white matter have been linked to specific object-naming deficits, suggesting a role for white matter tracts in semantic knowledge (Fang, Wang et al. 2018, Pestilli 2018). New methods are emerging that link cortical functional activity with white matter tracts (O'Muircheartaigh and Jbabdi 2018, Tarun, Behjat et al. 2020), shedding new light on how structural architecture might mediate macroscale dynamics or influence information flow. Furthermore, indices of white matter integrity have long been linked with clinical deficits, suggesting a possible role for white matter in disease models (Karlsgodt 2020, Hanekamp, Ćurčić-Blake et al. 2021). These studies suggest that the white matter has the potential to shape dynamics and impact cognitive processing.

The brain network model is in part useful because it abstracts the complex geometry and biology of the brain into a simple mathematical representation. When visualizing networks, often edges are represented as straight lines through space, with thicknesses or transparency that denotes edges strength. However, we should not lose sight that this representation is divergent from the anatomical reality of the brain, which is embedded in space and contains topographically organized white matter connections (Jbabdi, Sotiropoulos et al. 2015, Kurzawski, Mikellidou et al. 2020). Structural edges travel along physical paths through the white matter that have shape, curvature, and volume, and that compete for physical space and limited metabolic resources. Similarity of functional activity could be influenced by activity-dependent myelination (Fields, Woo et al. 2015), or possible ephaptic coupling of sheets of axons within white matter tracts (Sheheitli and Jirsa 2020). Thus, future work along these lines should focus on better understanding how the white matter plays a role in differentially shaping the relational content of brain networks.

## Subject-Specific Edge Information

Recent emphasis has been placed on extracting information from fMRI functional connectivity data, to characterize organizational features that robustly associate with a specific trait, like intelligence or attention (Finn, Shen et al. 2015, Rosenberg, Finn et al. 2016, Shen, Finn et al. 2017). This *connectome predictive modeling* approach involves filtering edges based on statistical criteria (such as correlation with a phenotype) and summing the edge weights for each subject. These sums are then used to create a statistical prediction model, in left-out subject data. The resultant cross-validated model outlines a set of edges important for predicting a desired phenotype. Notably, the networked characteristics of these edges can be analyzed to reveal system-level organization, such as the number of between system edges that participate in a high-attention predictive model (Rosenberg, Finn et al. 2016). This approach demonstrates the potential for mapping brain-behavior correlations at the level of brain edges. It remains to be seen how these predictive models could be extended to utilize edge constructs that capture higher-order relationships, which could be a productive future direction in tandem with the growing interest in applications of algebraic topology to brain network data.

## Conclusion

In contrast to brain network nodes, whose definition and differentiation have been the focus of brain mapping studies for years, issues and concepts relating to brain network edges have been less central to date. Here we have reviewed ways in which the edges matter, in terms of construction approaches that influence network analysis or in settings where groups of edges form higher-order relational information available for analysis. Furthermore, edges are a prime candidate through which to explore how communication processes unfold within the brain. Regardless of data modality, across neural data that spans spatial and time scales, we advocate for careful consideration of the information at the edge level. A greater focus on the information contained at the edges, otherwise known as an edge-centric perspective (de Reus, Saenger et al. 2014, Faskowitz, Esfahlani et al. 2020), can potentially stimulate novel exploration of brain organization. Both nodes and edges are fundamentally intertwined as the basic ingredients of a network model. Network neuroscience explorations can evidently benefit from both edge-centric and node-centric perspectives.

## Box: Alternative Weighting Strategies

Measurements of attributes that annotate existing edges can also be taken between neural elements (Fig. 1d). Edges of similarity and connectivity provide a quantification of the relationship between two nodes and collectively, form the topology of a given brain network. Already existing or estimated edges can be associated with metrics representing additional features, possibly derived from another modality or an embedding space. This approach allows for network edges to carry annotated layers of data derived from sources not directly related to the network construction process. Such features can aid computational modeling or data analysis. Attributes such as Euclidean distance, tract length, conduction delays, axonal caliber, biophysical efficacy, connection cost, or indices of myelination status are all examples of attributes that can be ascribed to edges expressing connectivity or similarity.

Edges can also be annotated with a value that reports a summary statistic or the result of combining several relational measures into a single weight. In this way, edges can carry weights that report a relationship generated from several modalities or conditions. Take morphometric similarity for example, which reports the correlation of standardized indices of myelination, gray matter, and curvature taken at the nodes (Seidlitz, Vasa et al. 2018). The edge weight here reflects a similarity across imaging domains that assess different aspects of the cortical geometry and composition. Within the realm of functional imaging, a generalized measure of functional co-activity between nodes can be estimated by combining data from rest and task sessions (Elliott, Knodt et al. 2019). Such a procedure can increase the reliability of intrinsic connectivity estimation. Relatedly, correlation values from various scan sessions can form a feature set at each edge, which can be used to create an edge-centric representation of edge covariance across conditions (Faskowitz, Tanner et al. 2021). Thus, edges can report multifaceted relationships incorporating a variety of data sources.

## Box: Parcellating Nodes or Edges

Even in a review of brain network edges, issues concerning the identification of nodes are worth noting. Edges are inexorably linked to nodes, documenting the relationship between the distinct elements of the neural system. The demarcation of neurons, neuronal populations, or cortical regions that constitute neural elements can be done using a range of methods (de Reus and Van den Heuvel 2013). A change in the definition of nodes will likely necessitate that the edges be recomputed. Early studies dividing the cortex based on neuronal tissue properties continue to influence present-day cortical mapping (Amunts and Zilles 2015). Other definitions of neural elements rely on the extraction of functionally coherent elements, such as the estimation of single units from an electrode array data (Dann, Michaels et al. 2016) or the grouping of spatial coherent and similarly active time series, ranging from the level of neurons to cortical vertices or voxels (Arslan, Ktena et al. 2018, Genon, Reid et al. 2018). Altogether, these methods describe how neural data can be parcellated, resulting in a set of nodes.

While the history of neuroscience is riddled with attempts to create nodal parcellations or maps of cortex (Finger 2001), considerably less attention has been devoted to defining or delineating distinct edges, for example tracts of the white matter. Commonly, features mapped in (cytoarchitectonics) or onto (connectivity) the cortex and subcortex are used as inputs for parcellation methods, which are essentially applications of node-based clustering and segmentation. However, it is also possible to cluster and segment data that relates directly to edges, specifically signals from the brain's white matter. For example, the streamline paths that result from tractography can be submitted to a hierarchical clustering routine, to create larger streamline groupings called bundles (Garyfallidis, Brett et al. 2012, Chandio, Risacher et al. 2020). Segmented tracts, when taken as fundamental building blocks of a network model, can be assembled into a matrix that records their intersections on cortical gray matter nodes. In such a model, tracts may be interpreted as conduits of specialized information or communication patterns that form elements of information processing (Pestilli 2018). In another example, bold-oxygen-level-dependent signal in the white matter can be clustered, forming parcels that relate to canonical systems found in the grey matter (Peer, Nitzan et al. 2017). These examples demonstrate alternative ways in which "edge" information could be conceptualized as neural elements. While little has been done so far, such an approach seems promising as it leads us to reconsider the primary importance of cortical nodes and may stimulate further modeling of organization found within the white matter.

## Box: Spatial Embedding Makes Brain Networks Unique

Networks are models of interrelationships between a system's elements. In many systems, there is no inherent cost to forming a connection. Consider the World Wide Web (WWW), in which nodes and edges represent URLs and hyperlinks, respectively. The "cost" of adding a hyperlink from one URL to another is minimal in that it requires no material contribution and (apart from the physical energy associated with writing HTML code) entails no metabolic or energetic expense. The lack of any explicit cost is a direct result of the fact that the WWW is not embedded in a physical space. The human brain, in contrast, is embedded in Euclidean space where the axonal projections and white-matter tracts require material to be formed and energy to be maintained and used for signaling (Stiso and Bassett 2018). For physical systems like the brain, forming and maintaining a network is costly. From a network's perspective, these costs are felt at the level of edges, where material and metabolic costs depend on geometric characteristics of anatomical connections, e.g. their length and diameter (Rivera-Alba, Vitaladevuni et al. 2011).

Brain networks are organized to reduce their material and metabolic expenditures, preferring to form short-range (and therefore less costly) connections. This preference, in turn, shapes the organization of the network and induces architectural features. For instance, networks that depend strongly on spatial constraints are naturally more clustered and readily form modules, making it difficult from an algorithmic perspective to adjudicate between "true" modules and those that reflect the underlying spatial constraints (Samu, Seth et al. 2014, Rubinov 2016).

On the other hand, brain networks do not strictly minimize their cost, forming a small number of long-distance connections (Betzel, Avena-Koenigsberger et al. 2016, Roberts, Perry et al. 2016). Presumably, these connections confer a functional advantage to the brain, otherwise we would expect evolution to have replaced them with shorter (and less-costly) connections. What roles do these costly long-distance connections play? In binary networks, they form "shortcuts" that reduce the network's characteristic path length and enhance communication efficiency (Kaiser and Hilgetag 2006). They also link high-degree nodes to one another, forming a constellation of interconnected hub nodes known as a "rich club", which plays a key role in the integration of information from different systems (Zamora-Lopez, Zhou et al. 2010, van den Heuvel and Sporns 2011). In weighted networks, however, long-distance connections play a reduced role due to their proportionally weaker weights (in spatial networks, connection weight tends to decrease monotonically with length). What role might these connections play? Across phylogeny, long-distance connections are both highly specific and robust, forming multiple bridges between the same distant neighborhoods. Recent work has suggested that these connections introduce unique and dissimilar signals into those neighborhoods, enhancing functional diversity and promoting increasingly complex dynamics (Betzel and Bassett 2018).

## Acknowledgements

We acknowledge the following manuscripts from which figures were adapted: Fig. 2 - Liu, Zheng et al. (2020) (CC BY 4.0), Balsters, Zerbi et al. (2020) (CC BY 4.0), Betzel, Medaglia et al. (2018) (CC BY 4.0), Sizemore, Giusti et al. (2018) (CC BY 4.0); Fig. 4 - Faskowitz, Esfahlani et al. (2020).

This material is based on work supported by the National Science Foundation Graduate Research Fellowship under grant no. 1342962 (JF). This material is based upon work supported by the National Science Foundation under grant no. 2023985 (RFB, OS). This material is based upon work supported by the National Institutes of Health and National Institute of Mental Health under grant no. 5R01MH122957 (OS). This research was supported by the Indiana University College of Arts and Sciences Dissertation Research Fellowship (JF). This research was supported by the Indiana University Office of the Vice President for Research Emerging Area of Research Initiative, Learning: Brains, Machines and Children (RFB).